\documentstyle[11pt,newpasp,twoside,epsf]{article}
\def\edcomment#1{\iffalse\marginpar{\raggedright\sl#1\/}\else\relax\fi}
\reversemarginpar

\begin{document}
\title{The Origin of Molecular Cloud Turbulence and its role on
Determining the Star Formation Efficiency} 
 \author{Enrique V\'azquez-Semadeni, Javier Ballesteros-Paredes}
\affil{Instituto de Astronom\' \i a, UNAM, Campus Morelia,
Apdo. Postal 3-72, Morelia, Michoac\'an, M\'exico}
\author{Ralf Klessen}
\affil{Astrophysikalisches Institut Potsdam, An der Sternwarte 16, 14482
Potsdam, Germany} 

\begin{abstract}
We suggest that
molecular cloud (MC) turbulence is a consequence of the very process of
MC formation by collisions of larger-scale flows in the
diffuse atomic gas, which generate turbulence
in the accumulated gas through bending-mode instabilities. 
Turbulence is thus maintained for as long as the accumulation process lasts
($\sim$ several Myr). Assuming that supersonic turbulence in
MCs has the double role of preventing global collapse while promoting
the formation of smaller-scale structures by turbulent compression
(``turbulent fragmentation''), we then note the
following properties: a) Turbulent fragmentation necessarily deposits
progressively smaller fractions of the total mass in regions of
progressively smaller sizes, because the smaller structures are subsets
of the larger ones. b) The turbulent spectrum implies that
smaller scales have smaller velocity differences. Therefore, below some
scale, denoted $l_{\rm eq}$, the turbulent motions become subsonic. This 
is an energy {\it distribution} phenomenon, not a dissipative one.
On this basis, we propose that the star formation efficiency (SFE)
is determined by the fraction of the total mass that is
deposited in clumps with masses larger than $M_{\rm J}(l_{\rm eq})$, the 
Jeans mass at scale $l_{\rm eq}$, because subsonic turbulence cannot
promote any further subfragmentation. In this scenario, the SFE
should be a monotonically increasing function of the sonic
and turbulent equality scale, $l_{\rm eq}$. We present preliminary
numerical tests supporting this prediction, and thus the suggestion that 
(one of the) relevant parameter(s) is $l_{\rm eq}$, and compare with
previous proposals that the relevant parameter is the energy injection scale.
\end{abstract}

\section{Introduction}

Two of the main questions concerning molecular cloud (MC) structure and star
formation are a) what is the origin and supply of MC turbulence? and b)
what is the origin of the low efficiency of star formation? Indeed, MCs
are known to be turbulent, with motions that are supersonic at scales
$\ga 0.1$ pc (Zuckerman \& Evans 1974; Larson 1981). 
However, recent numerical simulations (Padoan \& Nordlund 1999;
Mac Low et al.\ 1998; Stone,Ostriker \& Gammie 1998; Mac Low
1999; see also Avila-Reese \& V\'azquez-Semadeni 2001 for the case of
the global ISM)
have suggested that 
strong MHD turbulence decays rapidly even in
the presence of strong magnetic fields. Thus, it appears that MC
turbulence must be continually driven during the entire lifetime of
the clouds. 
The
driving mechanism, however, is probably not restricted to the stellar
activity {\it 
internal} to the clouds, since clouds devoid of stars exhibit similar
turbulent properties as clouds with stars (e.g., McKee 1999 and
references therein). 

Concerning the star formation efficiency (SFE), 
it is well known that it is
low, on the order of a few percent with respect to the total cloud
mass.
Traditional explanations for this low
efficiency have been, in the case of low-mass, isolated star
formation, that the MC cores in which stars form are magnetically
supported (i.e., ``subcritical''),
so that collapse is delayed until the magnetic field diffuses out of the
core by ambipolar diffusion (see, e.g., Shu, Adams \& Lizano 1987)
However, there
exist several recent suggestions that all cores are critical or
supercritical (e.g., Nakano 1998; Hartmann,
Ballesteros-Paredes \& Bergin 2001; Bourke et al.\ 2001; Crutcher,
Heiles \& Troland 2002).
On the other hand, it is known
that turbulence can prevent global gravitational collapse of a cloud
when the energy 
injection scale is smaller than the Jeans length (L\'eorat, Passot \&
Pouquet 1990; Klessen, Heitsch \& Mac Low 2000, hereafter KHM00), while
promoting 
``fragmentation'', i.e., the formation of smaller-scale density
substructures that can possibly undergo local collapse (Sasao 1973; Tohline,
Bodenheimer \& Christodolou 1987; Elmegreen 1993; V\'azquez-Semadeni,
Passot \& Pouquet 1996; Padoan 1995; Padoan et al. 2001; KHM00). KHM00
have shown that the efficiency in numerical simulations (measured as the
fraction of mass in collapsed objects) is large when the energy
injection scale is larger than the Jeans length, and low otherwise.
However, 
it is possible, as we suggest in \S 2
that MC turbulence is part of a cascade coming from larger scales.
Thus, the description in terms of an energy injection scale smaller than 
the MC itself is probably not optimal for real MCs. Also, this shows
that the origin of MC turbulence and of the SFE are intimately related. 

In the present paper, we propose that
{\it MC turbulence
originates from the same process that forms the clouds} (\S 2), and sets an
upper limit to the SFE 
through the fraction of the total mass it deposits in regions with
sizes such that the turbulent velocity dispersion becomes
subsonic (so that no further subfragmentation can occur within them),
and masses larger than their Jeans mass (\S 3).


\section{Origin of molecular cloud turbulence} \label{sec:turb_ori}

In this section we suggest, without proof, a plausible scenario
for the production of MC turbulence as a byproduct of their formation
process. It has been proposed that MCs are formed by the
convergence of large-scale streams in the diffuse ISM (Elmegreen 1993;
Ballesteros-Paredes, V\'azquez-Semadeni \& Scalo 1999, hereafter
BVS99). 
Although the MCs formed in the global ISM simulations of BVS99 
were at the limit of the resolution, rendering it impossible to resolve
the internal structure of the resulting clouds, one can infer it
from other pieces of evidence. This cloud formation mechanism is
analogous to the formation of shock-bounded
slabs between convergent flows, a process which
is known to be subject to fragmentation through nonlinear
instabilities (Vishniac 1994). This process has been simulated
numerically by various groups (Hunter et al.\ 1986; Klein \& Woods
1998; Folini \& Walder 1998), showing that it generates fully
developed turbulence in the shocked slab. Therefore, one expects
turbulence to be maintained for the entire duration of the
stream collision that forms the MC. In this scenario, MCs {\it and}
their internal turbulence are the consequence of a {\it lossy} energy
cascade in compressible turbulent flows, in which at every scale a
fraction of the energy is dissipated directly by shocks and the
remainder is passed to smaller-scale structures (Kornreich \& Scalo 2000).
Also, since the resulting MC is
not confined to a closed box, contrary to the case of simulations of
isolated clouds, and the compressible turbulence
generated produces both excesses {\it and} defects in the
density field, part of the material of the cloud can
be dispersed during the process. Numerical experiments
aimed at investigating this process in detail are now
underway using adaptive-grid and SPH numerical techniques, and will be
reported elsewhere.

\section{Turbulent Control of the Star Formation Efficiency}
\label{sec:turb_SFE} 

Assuming that turbulence is maintained for the lifetime of the MC as
described above, then the spatial redistribution (``turbulent
fragmentation'') of both mass and the various energies it produces
implies that only a fraction of the mass
in the turbulent cloud can end up in collapsed structures, assuming
that the total turbulent+thermal (collectively
referred to as ``kinetic'') energy is comparable to its
self-gravitating energy. In this case, the kinetic energy provides
support against global collapse of the whole cloud, while
simultaneously it promotes the production of smaller-scale density
structures through turbulent compressions, with the process
repeating itself at subsequently smaller scales, {\it until a change of
regime occurs}. A natural scale for a transition regime is the scale at
which the turbulent velocity difference becomes equal to the sound
speed\footnote{Note that $l_{\rm eq}$ is just a characteristic scale for
this to occur, there being a distribution of clump sizes at which the
equality occurs.}, denoted $l_{\rm
eq}$. Below this scale, the turbulence becomes 
subsonic, and is incapable of producing further
subfragmentation. Moreover, at this scale, thermal energy becomes
dominant in the support of the substructures, and thus the Jeans
criterion becomes the relevant one for determining whether a given
density peak 
(a ``core'') can become gravitationally dominated and collapse. Given
the statistical distribution of core masses and sizes formed by the
turbulence, only a fraction of the structures of size $l_{\rm eq}$
will have masses larger than the Jeans mass at this scale.

Since the turbulent fragmentation deposits ever smaller
fractions of the total mass in ever smaller-size structures, the
present scenario implies that flows with smaller values of $l_{\rm
eq}$ must have smaller fractions of mass available for collapse. This
can be tested experimentally by setting up simulations with different
ratios of turbulent to thermal enery, so that they have different
values of $l_{\rm eq}$. To do this, we use the simulations of
non-magnetic isothermal turbulence of KHM00, which
include cases with the same total kinetic energy,
but forced at different scales, so that $l_{\rm eq}$ is different, and
simulations with different amounts of kinetic energy, but similar
$l_{\rm eq}$. In fig.\ 1 we show the turbulent
energy spectra of three of these simulations, and their SFE, measured
as the fraction of mass in collapsed objects as a function of
time. This figure shows that runs with similar values of $l_{\rm eq}$
have similar SFEs, irrespective of their total kinetic energy, and
vice versa.

It is worth comparing with the proposal of KHM00 that the relevant
parameter is the energy injection (or ``driving'') scale of the
turbulence, $l_{\rm d}$. This parameter
clearly affects $l_{\rm eq}$ because, for fixed total turbulent energy
(given by the integral of the energy spectrum over $k$), injecting the
energy at smaller scales forces a larger energy content per wavenumber
range at small scales (compare the spectra of runs A1 and A3). However,
we see that runs B1 and A3, which were forced at different scales,
manage to have similar SFE because they have similar $l_{\rm eq}$. We
conclude that $l_{\rm eq}$ is the main parameter of the turbulent energy 
spectrum controlling the SFE. 

\begin{figure}
\plottwo{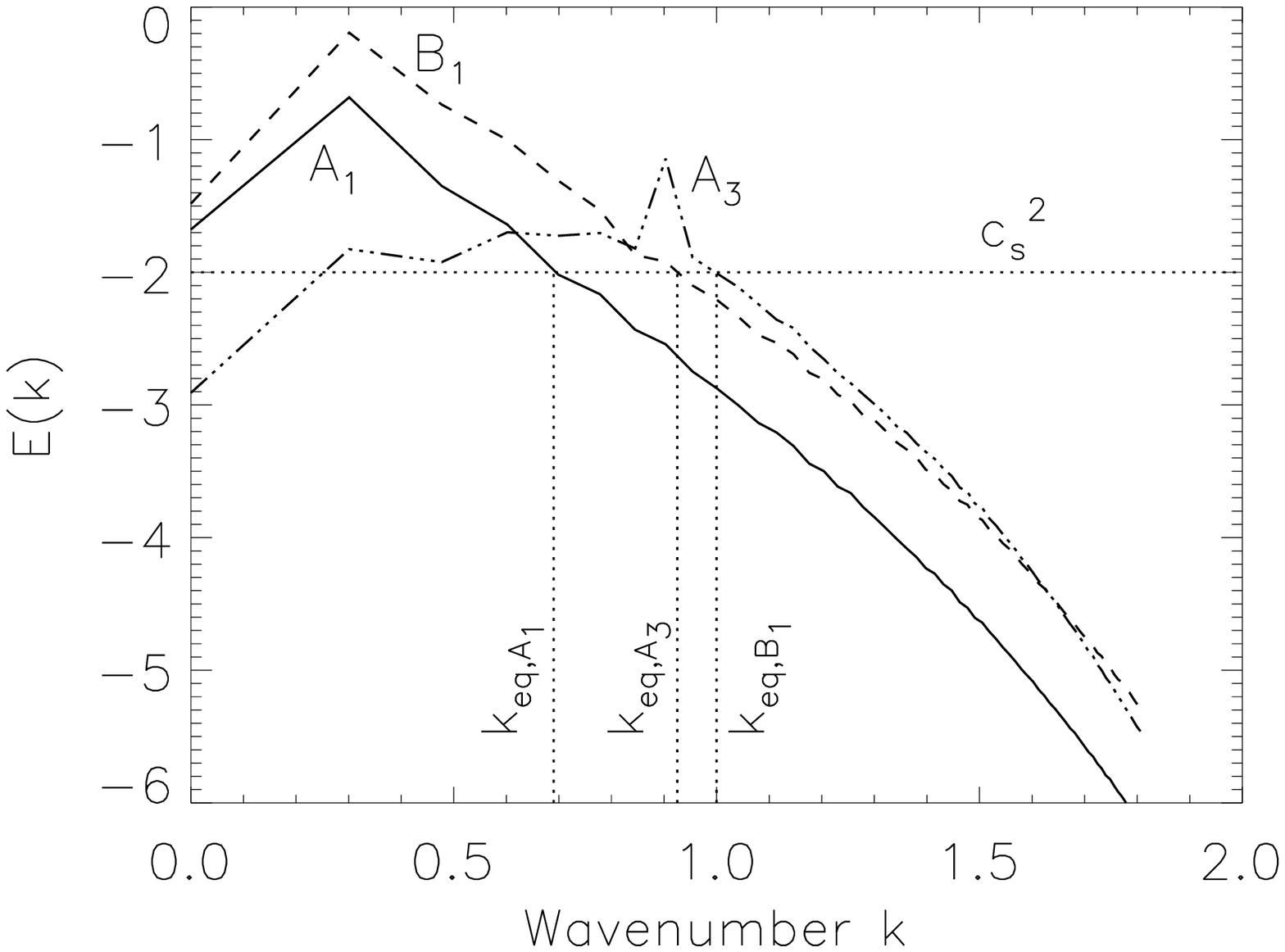}{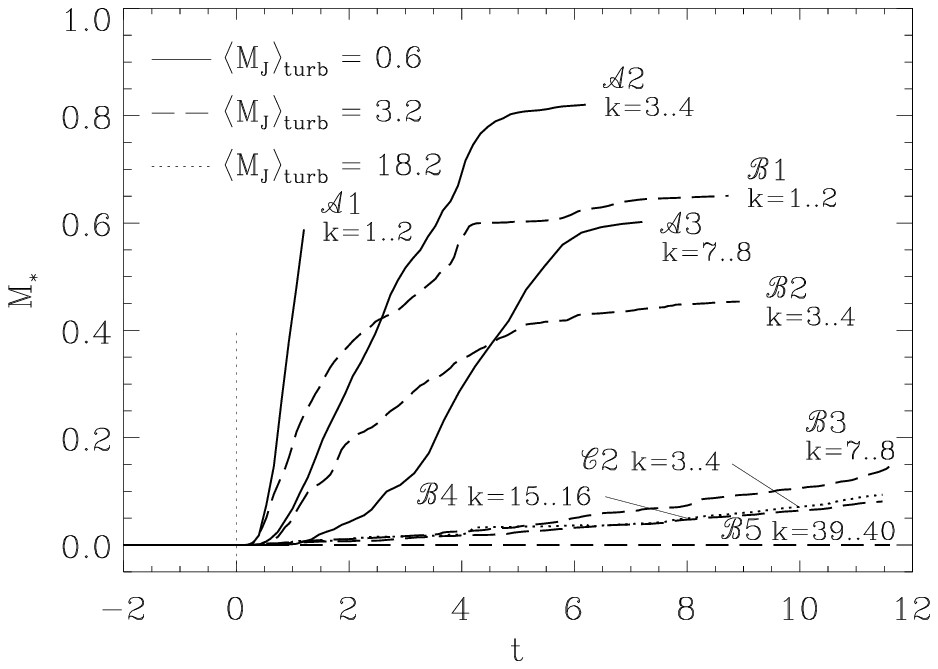}
\caption{({\it Left panel:}) Turbulent energy spectra for three
simulations of isothermal, 
self-gravitating, compressible, non-magnetic turbulence by KHM00, labeled
A1, A3 and B1, all with the same mass. Runs A1 and 
A3 have the same total kinetic (turbulent+thermal) energy, but different
spectral turbulent energy distribution, so that run A3 has more energy
at small scales (large wavenumbers $k$). Run B1 has more total kinetic
energy than A1 and A3, but its spectrum nearly coincides with that of A3
at small scales, thus having nearly equal values of $l_{\rm eq}$,
denoted by the vertical dashed lines. ({\it Right panel:}) SFE, measured 
as the fraction of mass in collapsed cores in the simulation as a
function of time (from KHM00). Runs A3 and B1 are seen to
have very similar values of the SFE.} 
\end{figure}

\section{Conclusions} \label{sec:discussion}

We have shown numerical evidence that $l_{\rm eq}$ is a relevant
parameter in setting an upper limit to the SFE (other processes, such as 
stellar energy feedback, may contribute as well). However, it is most
likely that other parameters are also involved. In particular, as the
spectral energy distribution of the turbulence varies, the sizes of the
density structures are likely to change as well. Padoan (1995) made an
attempt in this direction based on the probability distribution function
(PDF) of the density field. However, the PDF contains no information of
the mass spatial distribution. What is necessary is the density PDF {\it 
parameterized by region size}. This is precisely the kind of information
contained in the multifractal spectrum, and which we are in the process of
incorporating into the treatment.

\acknowledgements
We acknowledge CONACYT grant 27752-E to E.V.S.

\end{document}